\documentclass[conference]{IEEEtran}
\IEEEoverridecommandlockouts
\usepackage{cite}
\usepackage{amsmath,amssymb,amsfonts}
\usepackage{algorithmic}
\usepackage{graphicx}
\usepackage{textcomp}
\usepackage{xcolor}
\def\BibTeX{{\rm B\kern-.05em{\sc i\kern-.025em b}\kern-.08em
    T\kern-.1667em\lower.7ex\hbox{E}\kern-.125emX}}

\usepackage{enumitem}
\definecolor{gri}{RGB}{240, 242, 242}
\usepackage{adjustbox}
\usepackage{listings}
\usepackage{pgfplots}
\usepackage{array,multirow,graphicx}
\usepackage{float}
\usepackage{xcolor,colortbl}
\usepackage{subcaption}
\usepackage{todonotes}
\usepackage{listings}
\usepackage{scalerel}
\usepackage{tabularx, booktabs, makecell, caption}
\usepackage{xparse}
\usepackage{balance}
\usepackage{url}

\NewDocumentCommand{\ShowInline}{v}{%
#1%
}
\AtBeginDocument{%
  \providecommand\BibTeX{{%
    \normalfont B\kern-0.5em{\scshape i\kern-0.25em b}\kern-0.8em\TeX}}}

\begin{document}

\title{How Do Developers Use Code Suggestions in Pull Request Reviews?}

\author{\IEEEauthorblockN{Abir Bouraffa, Yen Dieu Pham and Walid Maalej}
        \IEEEauthorblockA{\textit{Department of Informatics} \\
        \textit{University of Hamburg}\\
        Hamburg, Germany \\
        {\{abir.bouraffa, walid.maalej\}@uni-hamburg.de}
}
}

\maketitle

\begin{abstract}
%\textbf{\textit{Background:}} 
 GitHub introduced the suggestion feature to enable reviewers to explicitly suggest code modifications in pull requests.
These suggestions make the reviewers' feedback more actionable for the submitters and represent a valuable knowledge for newcomers.
Still, little is known about how code review suggestions are used by developers, what impact they have on pull requests, and how they are influenced by social coding dynamics.
%\textbf{\textit{Aims:}} 
To bridge this knowledge gap, we conducted an empirical study on pull requests from 46 engineered GitHub projects, in which developers used code review suggestions.
%\textit{\textbf{Method:}} 
We applied an open coding approach to uncover the types of suggestions and their usage frequency.
We also mined pull request characteristics and assessed the impact of using suggestions on merge rate, resolution time, and code complexity.
Furthermore, we conducted a survey with contributors of the studied projects to gain insights about the influence of social factors on the usage and acceptance of code review suggestions.
%\textbf{\textit{Results:}} 
We were able to uncover four suggestion types: code style suggestions, improvements, fixes, and documentation with improvements being the most frequent.
We found that the use of suggestions positively affects the merge rate of pull requests but significantly increases resolution time without leading to a decrease in code complexity.
Our survey results show that suggestions are more likely to be used by reviewers when the submitter is a newcomer. 
The results also show that developers mostly search suggestions when tracking rationale or looking for code examples.
%\textbf{\textit{Conclusion:}} 
Our work provides insights on the usage of code suggestions and their potential as a knowledge sharing tool.
\end{abstract}


\maketitle

\section{Introduction}
% Significance/ Importance of the Topic
Code reviews have become an incontestable pillar of the software development process.
Beyond quality assurance, research has repeatedly shown ~\cite{GoogleCodeReviewSSCSB18, BosuCodeReviewOSSMicrosoft} that developers also use code reviews to share knowledge \cite{Dutoit:Book:2006, Maalej:TOSEM:2014}.
This can, e.g., be observed in pull request discussions on GitHub, where submitted code is reviewed, debated, and refined before a decision is made on whether or not to merge it into the main repository.
Software researchers and tool vendors have therefore worked toward facilitating such review discussions with the goal of improving the feedback process and the overall collaboration and knowledge sharing experience \cite{Maalej:PAKM:2008}.

\begin{figure}[h]
    \includegraphics[scale=0.42]{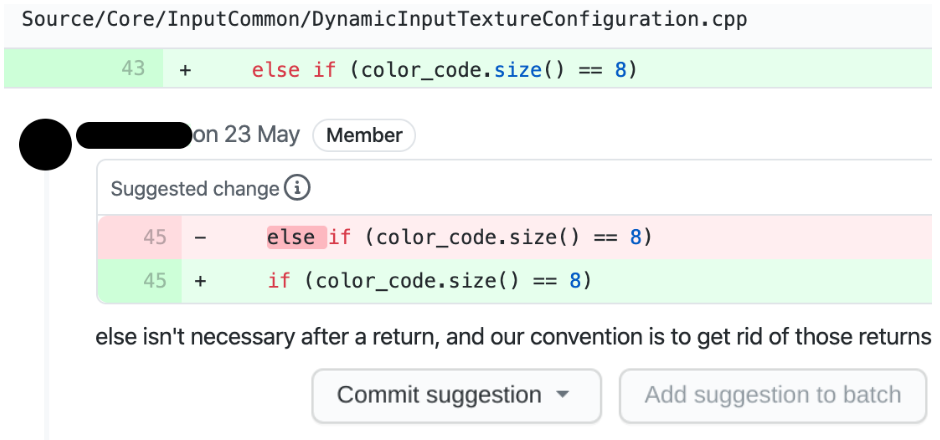}
    \caption{Example of a code suggestion in a GitHub pull request.}
    \label{fig:suggestion}
\end{figure}

In this spirit, GitHub introduced a specific structured review feedback feature called \textbf{code suggestions} in 2018~\cite{githubSuggestedChanges}, allowing reviewers to make direct changes to the pull request code, which could be committed immediately after.
Up until then, code changes suggested by the reviewer could only be written as comments, which were then interpreted and implemented by the submitter, only to be checked by the reviewer for correctness.
A suggested change can undergo one or more such iterations should the intent of the reviewer be misunderstood.
In contrast, code review suggestions, depicted in Figure \ref{fig:suggestion}, have the advantage of making the reviewer's feedback more structured and actionable by letting the reviewer implement their suggestion directly in the source code and enabling the submitter to integrate it with a single or bundled commit.
% Problem/ Knowledge Gap
% - Locate a Gap in the research, describe the problem you want to address, present a prediction to be tested etc.
% As an integral part of pull request discussions, code review suggestions are at the heart of these dynamics, which is why we  deemed it crucial to  study their influence.
Code review suggestions also make project best practices explicit through concrete code examples, which may provide guidance for future contributors lacking project experience \cite{Steinmacher:ITS:2015, Stanik:ICSME:18}.
Furthermore, code review suggestions can be leveraged in the future to offer support for the reviewers, for example by automating recurrent suggestions.

Despite its potential benefits, the usage and impact of the suggestion feature remain largely unexplored. 
Existing studies, such as the work by Brown and Parnin \cite{Brown_Suggestions_2020}, have provided insights into suggestion practices. 
However, the landscape has evolved since then, with broader adoption of the feature and emerging trends in automation and mentoring. 
This study seeks to fill this gap by investigating the types of suggestions, their perceived value, and their impact on pull requests in open source  projects.
% Owing to these potential benefits, code review suggestions represent an important avenue of research.
% The results can also be leveraged to further optimise the code review process by, for example, automating frequently suggested changes thus reducing the code review effort. 
% To better understand how code review suggestions are used in open source projects on GitHub, 
To achieve this, we mined a set of 46 repositories and surveyed the contributors of the corresponding projects. 
Our empirical investigation focuses on four main aspects, namely: the content of code review suggestions, their usefulness according to developers, their impact on the affected pull requests, and the factors influencing their creation and acceptance. 
These aspects are mirrored in our research questions defined as follows:

\begin{itemize}
    \item [\textbf{RQ1}] What types of suggestions do reviewers provide in pull requests? 
    \item [\textbf{RQ2}] What perception do pull request submitters and reviewers have of the different suggestion types?
    \item [\textbf{RQ3}] What impact do suggestions have on pull requests? %(answered by mining & survey --> roles)
    \item [\textbf{RQ4}] Which factors influence the creation and acceptance of code review suggestions? %(answered by mining & survey --> roles)
\end{itemize}

% Reminder of the paper
% The remainder of the paper is structured as follows. 
% Section \ref{sec:methodology} delineates the different analysis methods we used to answer our research questions.
% Section \ref{sec:results} describes the results of our analyses.
% Section \ref{sec:discussion} provides an interpretation of the results as well as their implications for future work. 
% Section \ref{threasts} relates the threats to the validity of the present work.
% Section \ref{sec:relwork} outlines related work and finally Section \ref{sec:conclusion} concludes the paper.

%--------------------------SECTION---------------------------

\section{Methodology}\label{sec:methodology}
We used a mixed-methods design for our empirical investigation of code review suggestions.
We mined suggestions from the pull requests of 46 GitHub repositories, which we use throughout our analysis as a reference dataset.
We then conducted a survey with the contributors of the studied projects to compare our results to the perception of developers (both pull request submitters and reviewers) with regard to the usage and acceptance of code suggestions.

%__________________SUB-SECTION_________________

\subsection{Mining Code Review Suggestions} \label{sec:Mining_Code_Suggestions}

%_____SUB-SUB-SECTION___

\subsubsection{Data Collection}
\label{sec:data_collection}
% Collection Process
Multiple initiatives have set the goal of providing a historical archive of GitHub repositories.
Among the most widely known and impactful works is GHTorrent, which was initiated by Gousios~\cite{GHTorrent_Gousi13} and spans over 18TB of compressed metadata~\cite{ghtorrent}.
However, not all repositories on GitHub are fit to be considered for research.
Munaiah et al.~\cite{Repo_Dataset_Munaiah_2017} have presented tools capable of separating the so-called \textit{signal}, i.e.~engineered software repositories, from the \textit{noise} such as educational assignments.
To test their classification approach, Munaiah et al.~manually annotated a sample of 300 GitHub repositories, of which half were considered as engineered software repositories.
We selected this initial set of 150 repositories, which we then subjected to filtering based on the following criteria inspired by Levin et al.~\cite{Commit_Classification_Levin_2017}:
\begin{itemize}
    \item Has more than 100 stars
    \item Has more than 60 forks 
    \item Updated since 01-01-2020 
    \item Exists for more than 3 years
    \item Exceeds 2MB in size
\end{itemize}
% Total number of pull request and further information about the pull request, like e.g. number of commits, deletions etc.
This resulted in a set of 95 repositories.
As a next step, we queried pull requests and discussions created since feature introduction on October 16 of 2018  using the PyGithub API~\cite{pygithub}.
Since GitHub suggestions are enclosed between  \texttt{```suggestion} and \texttt{```}, we used regular expressions to parse the review comments of each pull request and extract suggestions.
The resulting set of 46 repositories includes 2,852 pull requests using suggestions and a total of 8,672 suggestions.
From the 46 repositories, we additionally extracted 24,150 pull requests created after feature introduction where suggestions were not used.
These pull requests were used in our impact analysis to compare the characteristics of pull requests using the suggestion feature to those of pull requests not using it.
% For our regression analysis assessing the impact of suggestions on pull requests
We share the list of studied repositories and their characteristics on Zenodo~\cite{zenodoRepositorySample}, and the full replication package for this study on GitHub\footnote{\url{https://github.com/abiUni/chase25_replication-package}}.

All pull requests were then labelled as ``merged'' or ``not merged'' based on the content of the field \textit{merged\_at} and as ``with suggestions'' or ``without suggestions''.
% We should note that we only consider pull requests having a \textit{closed} state.
If the pull request was merged, the time to resolution was calculated in hours as the difference between the merge date and the creation date. Otherwise the resolution time was calculated as the difference between the creation and closure dates.
We determined the association of the submitter of the pull request with the project through the \textit{author\_association} field.
This field is also provided in review comments, allowing us to determine the reviewer's association with the project.
The developers in our sample, either the submitter or the reviewer, have one of the following project associations~\cite{githubEnumsGitHub} reflecting their experience in the project.
\begin{itemize}[leftmargin=*]
    \item{\textbf{Member:} member of the organization owning the repository}
    \item{\textbf{Collaborator:} user was invited to collaborate}
    \item{\textbf{Contributor:} user has committed to the repository}
    \item{\textbf{None:} user has no association with the repository}
\end{itemize}

% %_____SUB-SUB-SECTION____

% \subsubsection{Data Preparation}
% % Defin Data Sets for the different analysis procedures

% %_____SUB-SUB-SECTION___

\subsubsection{Open Coding}
\label{sec:open_coding}
%Goal
To answer RQ1 and uncover code review suggestion types, we applied an open coding approach.
%Data Set/ Preparation
We focused our analysis on a \textit{subset} of the total of 8,672 code review comments with suggestions.
We only considered review comments in which the  suggestion was accompanied by a natural language comment to reduce interpretation errors.
Moreover, we only considered natural language comments written in English and filtered out comments not deemed meaningful for coding (e.g.~only containing emojis, a reference to a commit link, or a statement like ``Resolved'' or ``Wontfix'').
The resulting set consisted of 2,332 review comments representing 26.9\% of our total sample.

%Process
The peer coding was done by the first author and a student research assistant. 
Care was taken into randomising the order for each of the coders to prevent possible biases.
In the first phase of our analysis, we independently coded a random sample of 100 code review comments.
We then discussed the codes and agreed on an initial coding guide, which we incrementally reviewed and enriched during the following phases.
In the second phase, we proceeded to code further 1000 randomly selected code review comments.
During the second phase, no new codes were discovered and we proceeded with the final coding round of the remaining 1,232 items.
This last step did not result in additional codes either.
With the coding complete, we analysed our inter-rater agreement score and reviewed a subset of mismatches to better understand the sources of conflict.
Our percentage agreement came to 76.1\% corresponding to 1,775 matches and a Cohen's Kappa of 0.65, which is considered a substantial agreement \cite{landis_measurement_1977}.

%_____SUB-SUB-SECTION____

\subsubsection{Cyclomatic and Cognitive Complexity}
%Goal
To answer RQ3 and better understand the effects of code suggestions on the complexity of the affected code, we compared the cyclomatic complexity (CC) and number of code lines (LOC) before and after the introduction of code suggestions to the source code.
%Data Preparation/Set
We were able to analyse the complexity on the function level of 2,690 source code files from six programming languages, to which a code suggestion was made.
We should note that we do not recover the commit integration of the code suggestion by the pull request author, since code suggestions can be  bundled with other changes into a single commit.
Instead, we constructed the modified code version by integrating the code suggestion into a copy of the original commit.
We achieved this using the \textit{original\_position} field in the review comments, which indicates to which line in the original commit the suggestion was made.
%Analysis Process
To analyse Mccabe's Cyclomatic Complexity (CC)\cite{McCabe_Complexity_2016}, we used an open source analyser called \textit{lizard}~\cite{githubGitHubTerryyinlizard}, which allowed us to extract CC and LOC at the function level.
We were also able to report the change in cognitive complexity\cite{campbell2018cognitive} for 2,107 of the functions. 
This metric better reflects the mental effort needed to understand code by evaluating factors like nested logic, control flow, and readability. 
It focuses on human cognitive load and has been shown to correlate positively with comprehension time and subjective ratings of understandability \cite{munoz2020empirical}.

%%%%%%%%%%%%%%%%%%%%%%%%%
%%%%%%%%%%%%%%%%%%%%%%%%%

%_____SUB-SUB-SECTION____
\begin{table}[t]
    % \centering
    \renewcommand{\arraystretch}{1.1}
    \setlength{\tabcolsep}{9pt}
    \caption{Familiarity with the code suggestion feature.}

    \begin{tabular}{l r}
    \toprule
\textbf{Are you familiar with the CR suggestion feature?} & \textbf{\#}\\
    \midrule
    No  &  30\\
     \multirow{2}{*}{\makecell[l]{Yes, I have seen it in pull requests but I have not used \\or received any code review suggestions}}  & 13\\
     &\\
     Yes, I have received CR suggestions on previous pull requests& 18\\
     Yes, I have used CR suggestions in my reviews & 11\\
     Yes, I have received and used CR suggestions before & 50\\

     \midrule
     \textbf{Total} & \textbf{122}\\
    \bottomrule
    \end{tabular}
    
    \label{tab:survey_familiarity}
\end{table}

\subsubsection{Multiple Regression Analysis}
%Goal
To answer RQ3 (i.e.~determine what impact the use of suggestions has on pull requests' merge rate and resolution time) we conducted a multiple regression analysis.
%Data Set/ Preparation
We only considered  pull requests which had the status \textit{``closed''}.
This resulted in a dataset of 24,150 pull requests \textit{without} suggestions of which 15,867 were merged (merge rate: 65.7\%) and 2,548 pull requests \textit{with} suggestions of which 1,942 were merged (merge rate: 76.2\%).
We chose multiple regression as a statistical instrument because it allows us to take into account multiple independent variables \cite{creswell_research_2018}, representing the following pull request characteristics:
\begin{itemize}[leftmargin=*]
    \item \textbf{with suggestions}: binary indicator of suggestion usage ($x_1$)
    \item \textbf{additions}: number of lines added ($x_2$)
    \item \textbf{deletions}: number of lines deleted ($x_3$)
    \item \textbf{changed files}: number of files changed ($x_4$)
    \item \textbf{commits}: number of commits ($x_5$)
    \item \textbf{comments}: total number of comments ($x_6$)
    \item \textbf{review comments}: number of review comments ($x_7$)
    \item \textbf{submitter role}: submitter's association with the project ($x_8$)
\end{itemize}
%Process
The general regression equation for each of our dependent variables, i.e.~merge outcome and resolution time in hours generally denoted by $\hat{y}$, and taking into account the error term $\epsilon$ reads as follows:
\begin{subequations}
  \begin{align}
&\hat{y}=\beta_{0} + \beta_{1}*x_{1}
+ \dots + \beta_{8}*x_{8} + \epsilon 
\end{align}
\end{subequations}
Accordingly, our null hypotheses state that $x_1$ does not affect the dependent variable approximated by the regression:
\begin{subequations}
  \begin{align}
&\pmb{H_{0}}: \beta_{1} = 0,\quad\pmb{H_{1}}: \beta_{1}\neq 0,\quad \alpha =0.05
\end{align}
\end{subequations}

\subsubsection{Suggestion Use and Project Association}
%Goal
Pull request discussions are influenced by various factors, many of which are social in nature as outlined by Tsay et al.~\cite{Tsay_Discussions_GitHub_2014}.
% The authors report discussion dynamics whereby reviewers' reaction to a pull request ranged from a disapproval and questioning the solution approach to proposing an alternative solutions or even advertising their own solution.
Moreover, the authors pointed out that the submitter's prior project experience changed the nature of the discussion around their pull request contribution.
To uncover possible social dynamics influencing the use and acceptance of code review suggestions and answer RQ4, we analysed the distribution of the suggestion types discussed in  Section \ref{sec:open_coding} with regard to the submitter's and reviewer's association with the project as defined in Section \ref{sec:data_collection}.

%__________________SUB-SECTION_________________

\subsection{Survey}\label{sec:survey}
\label{sec:survey_prep}
We followed up our data analysis with a survey of developers from our project dataset, to gain insights about their perception of code review suggestions.
The survey partly serves as a triangulation of the data mining analysis and partly adds new results answering R2-RQ4.

%_____SUB-SUB-SECTION____

\subsubsection{Design}
% , mainly on whether the feature is used for other purposes and whether the project association has an influence on the usage and acceptance of suggestions.
The survey includes both closed and open-ended questions and five-point Likert Scale questions with agreement from strongly disagree (1) to strongly agree (5).
We structured the survey in three main parts.
In the first part, we asked the participants general questions about their open-source project experience.
% and the activities they engage in within those projects.
% We also asked questions about the goals of the code review process in place.
In the second, we asked about the participants' experience with the GitHub suggestion feature during code reviews.
We asked two Likert scale questions about their perception and usage of suggestions both from the perspective of a pull request submitter and a reviewer.
% In order to triangulate the findings from the code review mining in section \ref{sec:Mining_Code_Suggestions} with the actual perception of the project developers, 
To differentiate perspectives, we conditioned the questions on the level of familiarity with the suggestion feature, which the participants had to answer first.
We defined the familiarity levels as depicted in Table \ref{tab:survey_familiarity}.
The second section also contained questions about secondary uses of code review suggestions as an information source.
Lastly, we gathered information about the participants' demographics in a third section.
The average participation time was 15 minutes. 
The survey was configured to ensure anonymity using LimeSurvey.
\begin{table}
\centering
        \caption{Demographics of the survey participants.}
    \begin{tabular}{l rrrr}
    \toprule
    \textbf{Attribute} & \textbf{min} & \textbf{max} & \textbf{mean} & \textbf{median} \\
    \midrule
    \textbf{Age} & 20 & 78 & 34.5 & 34\\
    \textbf{Years in OSS} & 2 & 30 & 10 & 8 \\
    \midrule
    & \textbf{Male} & \textbf{Female} & \textbf{ Non-binary} & \textbf{Undisclosed} \\
    \midrule
    \textbf{Gender} & 112 & 3 & 1 & 5\\
    \midrule
    & \textbf{None} & \textbf{Member} & \textbf{Contributor} & \textbf{Collaborator} \\
    \midrule
    \textbf{Roles} & 17 & 97 & 113 & 79\\
    \bottomrule
    \end{tabular}

    \label{tab:demographics}
\end{table}
\begin{table*}[ht]
\centering
    \caption{Types of suggestions in pull requests reviews and their prevalence in our dataset.}
    \begin{tabularx}{0.85\textwidth}{l r r l}
    \toprule
     \textbf{Type} &  \textbf{Count} & \textbf{\%} & \textbf{Example}\\
     \midrule
        Improvement & 861 &48.51\% & \parbox[c]{1em}{
      \includegraphics[scale=0.35]{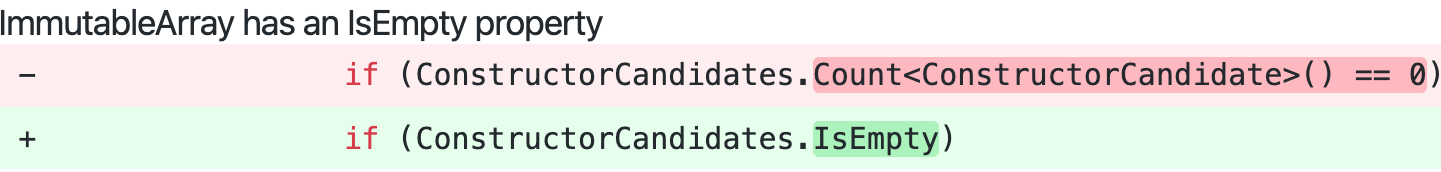}}\\
      \midrule
        Documentation &  468 & 26.37\% & \parbox[c]{1em}{
      \includegraphics[scale=0.35]{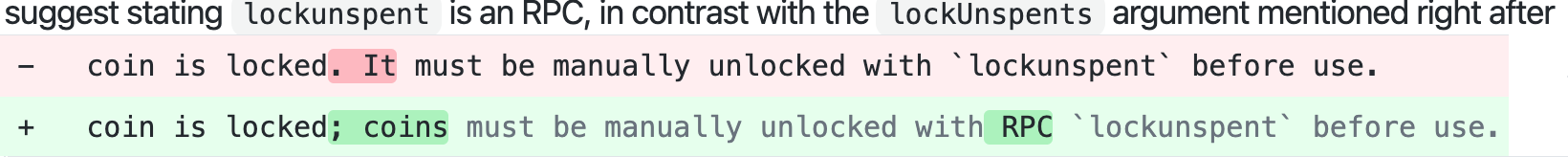}}\\
      \midrule
        Code Style & 304 & 17.13\% & \parbox[c]{1em}{
      \includegraphics[scale=0.35]{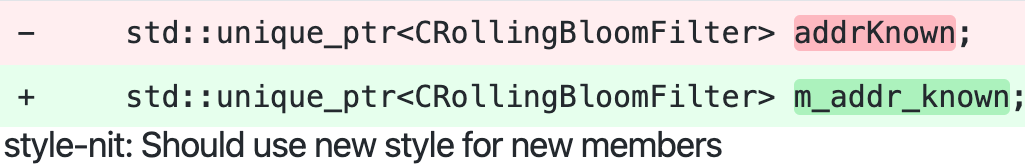}} \\
      \midrule
        Fix & 142 & 8.00\% & \parbox[c]{1em}{
      \includegraphics[scale=0.35]{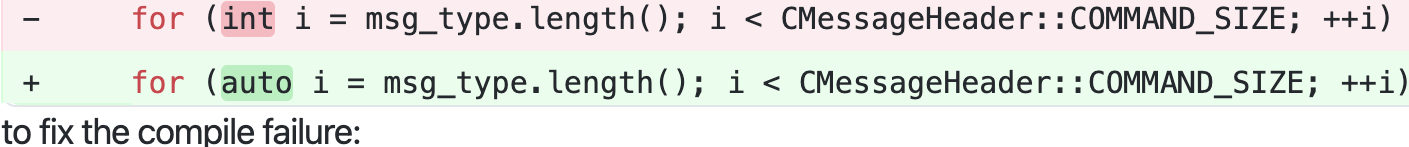}}\\
      \midrule
        \textbf{Total} & \textbf{1,775} & \textbf{100\%} & \textbf{-} \\
    \bottomrule
    \end{tabularx}

    \label{tab:suggestion_types}
\end{table*}

%_____SUB-SUB-SECTION____

\subsubsection{Procedure}
We started the first round of distribution of email invitations on April 27, 2021 and followed up with a second round on May 04.
The survey was closed a week later on May 10.
We provided information about the purpose of the study, the target group, privacy policy, and asked for the participants' consent prior to taking part in the survey.
Our first round of email invitations went to 445 users representing the pull request authors and code reviewers who either used or received suggestions based on our sample dataset of code review suggestions from section \ref{sec:Mining_Code_Suggestions}. 
We then expanded our survey to all the contributors of the studied projects eliminating the first target group to avoid spamming.
The second target group consisted of 3,326 users.
We only wrote to users who disclosed their first name and email address on their public profile signalling their interest to be contacted.
We required all participants to be at least 18 years old.

%_____SUB-SUB-SECTION____

\subsubsection{Participants}
Overall, we contacted 3,771 GitHub users and received 122 survey participations corresponding to a response rate of 3.24\%.
 From the first round with 445 users,  we received 26 responses corresponding to a response rate of 5.84\%.
From the second round with 3326 users, we received 96 responses corresponding to a response rate of 2.56\%.
Despite requiring that the respondents be 18 years or older, one respondent stated being 15, which led us to discard their participation, bringing the number of responses to 121. 
As outlined by Table \ref{tab:demographics}, the participants were predominantly male (92.6\%), had a median age of 34 and have been involved in open source projects for a median of 8 years.
The participants were also asked what roles they currently occupy or have occupied in the past.
To this questions, 93.4\% of participants replied that they have been or currently are a contributor, 65.3\% replied that they acted or are currently acting in the role of a collaborator and 80.2\% replied that they have been or are currently a member.
Only 17 stated having no role in a project, 3 users stated being repository owners and one person described their role as ``single maintainer''.

%_____SUB-SUB-SECTION____

\subsubsection{Pilot}
To test its comprehensibility, we ran the survey with members of our research group who have experience working in open source projects and who use GitHub. 
The pilot comprised three iterations, each with a different set of researchers. 
Based on the results, we refined the questions, reordered them to reduce learning bias and added definitions for the different developer roles as stated by GitHub. 
% The feedback also helped us identify improvements, adjust answer choices and change the wording and the order of some questions.
We also reduced the number of questions and time-boxed the survey to 15 minutes, on average.
%--------------------------SECTION---------------------------

\section{Results}\label{sec:results}

%__________________SUB-SECTION_________________

\subsection{RQ1: Suggestion Types}
\label{sec:result_types}
Following the open coding analysis, we uncovered four suggestion types: \textit{fix}, \textit{code style}, \textit{documentation}, and \textit{improvement}. 
In the following, we define each type, how it is motivated by the reviewers in the comment as well as the frequency with which it occurred.
Table \ref{tab:suggestion_types} summarises the prevalence of the different types and provides examples for each.
% We discovered a large variety of motivations that reviewers mentioned for suggesting changes, which serve both as an explanation and a justification.

%_____SUB-SUB-SECTION____

\subsubsection{Code Improvement}
%Frequency 
The most prevalent type of suggestions in our sample are improvements, which cover almost half of all suggestions at 48.51\%.
%Definition
An improvement consists of a suggestion affecting the source code excluding code comments and  mostly aims to ameliorate the code in a number of possible ways: either by reducing complexity, increasing performance, or improving maintainability.
%Explanation/ Justification
One motivation that often co-occurred with improvement suggestions is simplification, whereby the reviewer points to an unnecessary complexity, a redundancy, or verbosity prompting the improvement suggestion.
For example: \textit{``I must say things seem a bit overcomplicated [...] Any reason not to make it simpler?''}~\cite{githubH2ClientProtocolAdityaa30}.
Understandability also motivated this type of suggestions, meant to increase code comprehension as:
\textit{``Use an equivalence key that indicates the operation being applied.''}~\cite{githubUpdateGenerateHashCode}.
Some reviewers motivated their improvement suggestion with performance optimisations, e.g.: \textit{``minor optimization (saves like 10 lines of assembly)[...]''}~\cite{githubWASAPICode} while others aimed at increasing maintainability: \textit{``Couldn't this be in a function, so that only one place has to be updated in the future?''}~\cite{githubRpcHuman}.

%_____SUB-SUB-SECTION____
\begin{figure*}[t]
\centering
\begin{subfigure}[b]{0.4\textwidth}
    \includegraphics[scale=0.35]{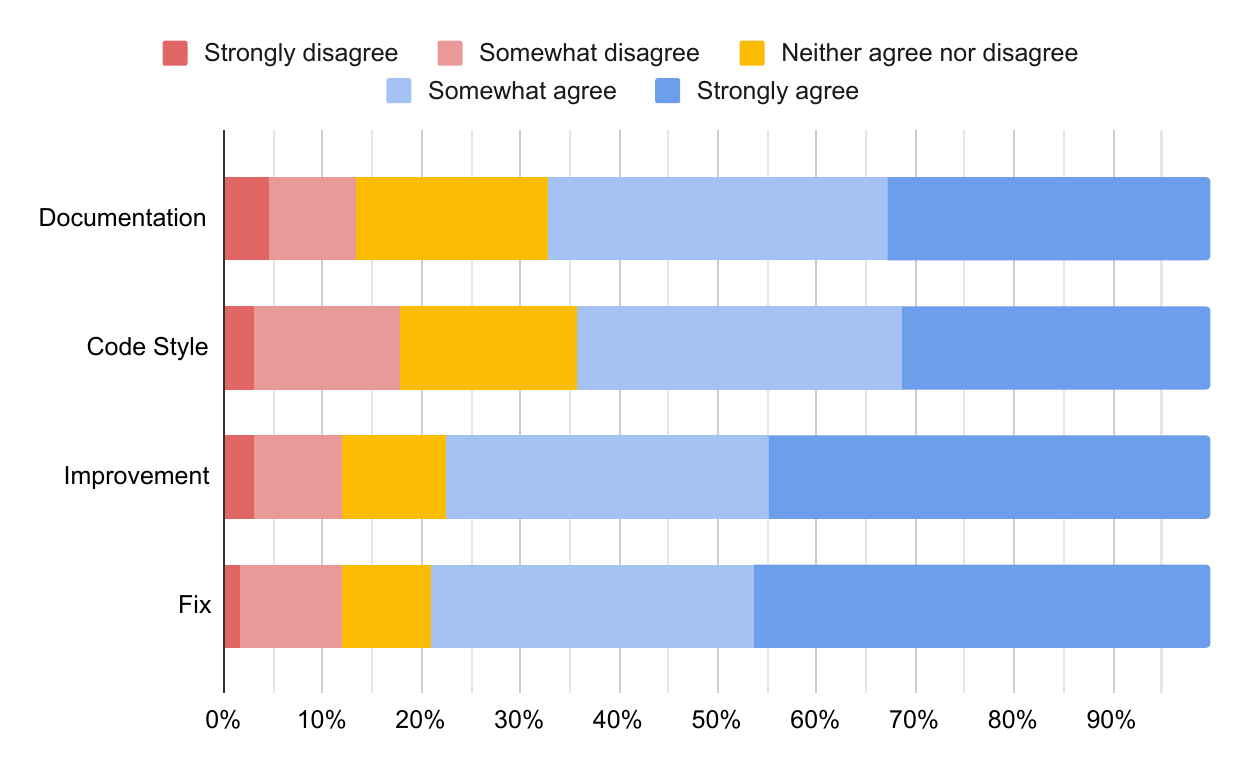}
    \caption{Submitter perception of suggestion type usefulness.}
    \label{fig:submitter_acceptance}
\end{subfigure}
\hspace{2em}
\begin{subfigure}[b]{0.4\textwidth}
    \includegraphics[scale=0.35]{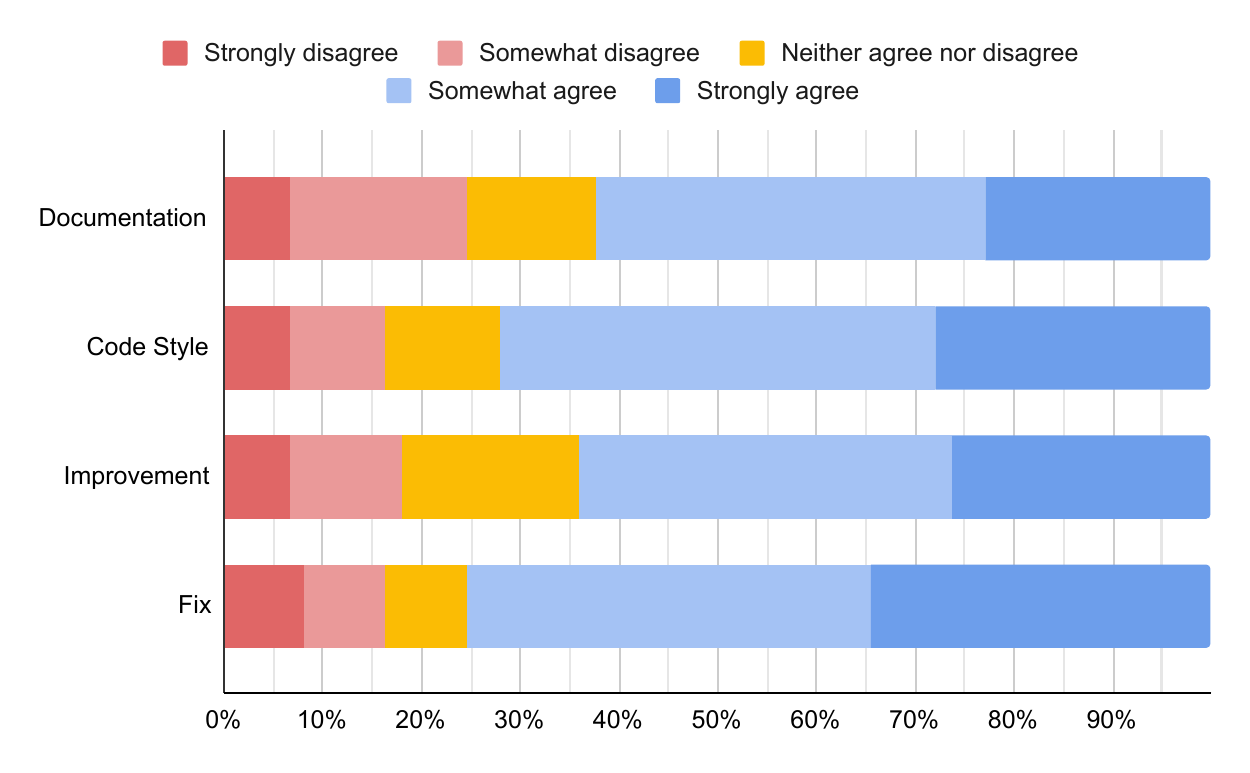}
    \caption{Reviewer usage of suggestion types.}
    \label{fig:reviewer_likelihood}
\end{subfigure}
\caption{Perception of suggestion types: pull request submitters vs.~reviewers.}
\label{fig:perception_suggestion_types}
\end{figure*}
\subsubsection{Documentation} 
%Frequency 
The second most frequent suggestion type in our sample is \textit{Documentation} with 26.37\%.
%Defintion
Documentation refers to code review suggestions which either modified official project documentation, for example a \textit{ReadMe.md} file, or code comments.
This category differs from other suggestion types, specifically from code style, in that it modifies  the content and not merely the structure or formatting.
This category was often used by reviewers to reformulate a confusing or wrong statement about code behaviour or to update outdated information such as a URL, a version number, or code examples. 
One developer, e.g., wrote: \textit{``I think mentioning ``component" is more likely to confuse than clarify here.''}~\cite{githubFLINK16210docsExtending}.

% These suggestions serve an important role in keeping the documentation in tune with the code and are often motivated by concerns about confusions that might arise from certain formulations as outlined in

%_____SUB-SUB-SECTION____

\subsubsection{Code Style} 
%Frequency
Code style suggestions are less frequent in comparison to the first two types making up only 17.13\% of the suggestions in our sample.
%Definition
Code style suggestions are code rewriting changes that modify the appearance of the code without affecting code functionality or behaviour.
While a considerable amount of reviewers mention their suggestions as a \textit{``code style"}, \textit{``nitpicking"} or \textit{``lipstick"}, we could not discern a clear and consistent definition of code style across projects. 
To counter this ambiguity, we consulted multiple code style guides including the google java style guide~\cite{googleGoogleJava} as well as research related to code style criteria such as the work of Zou et al.~\cite{Zou_Code_Style_Criteria_2019}.
We concluded from our search that code style is commonly defined through four criteria: \textit{code structure}, \textit{formatting}, \textit{naming} and \textit{commenting}. 
We consequently adopted this definition for our coding.
% \textit{Code structure} refers to the order and organisation of components within a package or file. 
% An simple example hereof are the orders in which imports are organised in the beginning of a file.
% \textit{Formatting} encompasses a variety of rules concerning spacing and indentation, bracketing, maximum length of code lines among other rules.
% \textit{Naming} is another criteria that aims to increase readability by setting different naming rules depending on component type and locality.
% Lastly, \textit{commenting} refers to the structure of code comments such as the use of certain tags.
% These type of suggestions are either to adhere to the project's code style, a general language-specific code style or stating the personal code style preference of the reviewer.

%Explanation/ Justification
Code style suggestions were in some cases motivated by compliance to the language-specific style guidelines or to the project conventions. 
An example of the latter is: \textit{``Also our current convention is to place standard includes after the corresponding header include for the cpp file (aka, the first one).''}~\cite{githubSupportWindows}
In some other cases, the reviewer simply expressed their personal code style preference as motivation for the suggestion, e.g.: \textit{``I personally prefer identifiers to be monospaced [...] Would you mind updating them here?''}~\cite{githubResourcePacks}
Other reviewers were concerned about code readability, writing: \textit{``the reason for this is that, by default, when a method is single-line it just looks abstract to me. and i use enough cognition to just scan the signature.''}~\cite{githubExtractClass}

%_____SUB-SUB-SECTION____

\subsubsection{Fix}
%Frequency
Suggestions to fix a defect in the code are the rarest with only 8\% of suggestions belonging to this type.
%Frequency
Code review suggestions of this type are mainly intended to fix a defect discovered by the reviewer in the submitted commit.
Suggestions of this type may also be of preventive nature by offering a solution to \textit{undesirable} implications or \textit{side effects}.
An example of this type of suggestions in our dataset is: \textit{``note: without this change, your fix is buggy and will crash in the following case: Foo(x == true), please add test.''}

\subsection{RQ2: Perception of Suggestion Types}
\label{sec:perception_suggestion_types}
Our survey included two Likert scale questions about the perception of suggestion types both from the perspective of pull request submitters (67 participants) and reviewers (60 participants).
The participants received either one of the two questions or both depending on their familiarity with the suggestion feature as outlined in Table \ref{tab:survey_familiarity}.

\subsubsection{Submitter Perspective} We asked pull request submitters who have received suggestions about how useful they perceive each suggestion type.
For example, the statement for code style suggestions read as follows: \textit{``I find code review suggestions about code style changes useful''}.
The respondents stated their agreement on a 5-point scale from \textit{strongly disagree} to \textit{strongly agree}.
The results are outlined in Figure \ref{fig:submitter_acceptance}.
Overall, we neither observed strong differences nor opinion-divides. 
We found that fix and improvement suggestions received the highest strong agreement scores of 45.6\% and 44.1\% while documentation and code style suggestions had lower agreement (32.8\% and 31.3\%) as well as the highest disagreement scores of 17.9\% and 13.9\% respectively.

\subsubsection{Reviewer Perspective} Since participants can have experience with suggestions both as submitters and as reviewers, we avoided asking about suggestion type usefulness for the reviewer perspective to avoid redundancy.
Instead, we asked participants of this group whether they use the different suggestion types in their reviews.
For example, the statement for suggestions of type fix read as follows: \textit{``I use code review suggestions in my reviews to provide solutions to defects in the code''}.
The respondents were able state their agreement on a 5-point range from \textit{strongly disagree} to \textit{strongly agree}.
The results of the this question are outlined in Figure \ref{fig:reviewer_likelihood}.
The overall agreement surpassed 60\% for each suggestion type.
Fix and code style suggestions both received strong agreement scores of 34.4\% and 27.9\% and slight agreement scores of 41\% and 44.3\%.
The highest rate of overall disagreement was recorded for documentation with 6.6\% of reviewers strongly disagreeing that they use this type of suggestion and 18\% only slightly disagreeing with the statement.

Overall, all suggestion types appear to be welcomed by submitters and in use by reviewers.
However, submitters consider fixes and improvements slightly more useful whereas reviewers stated using more fix and code style suggestions.
% We can conclude that a balance of supply and demand is given for fixes but this balance  is rather tilted for improvements. 

%__________________SUB-SECTION_________________
\begin{figure}[t]
\centering
    \includegraphics[scale=0.45]{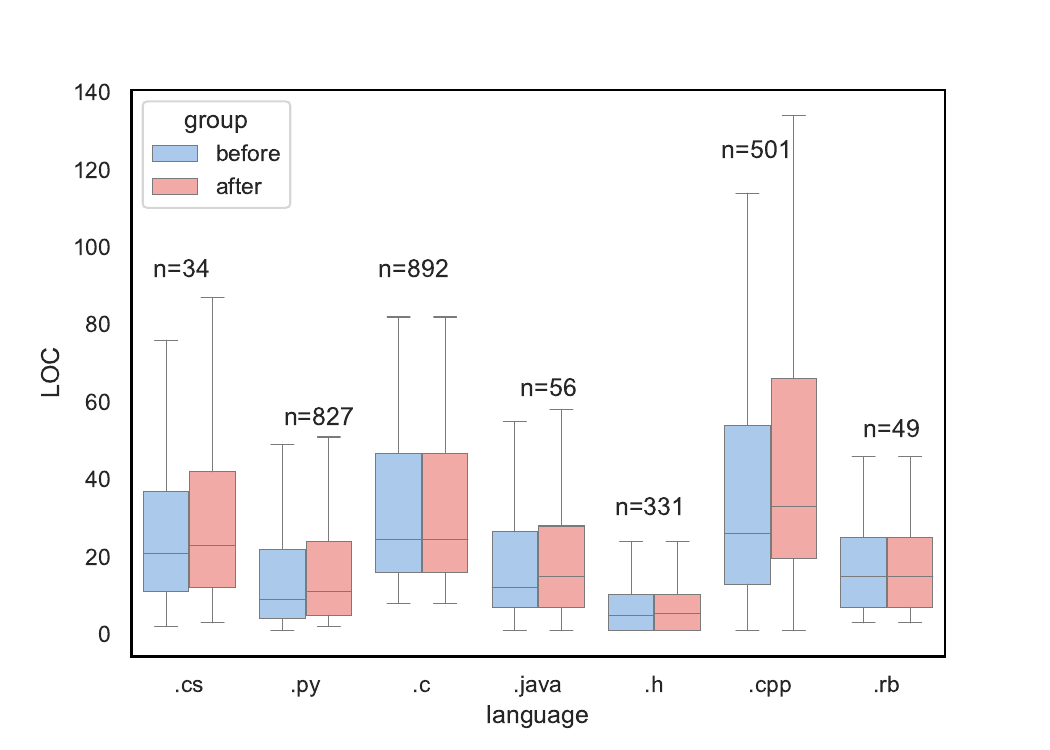}
    \caption{LOC before and after the suggestions.}
    \label{fig:fucntion_LOC}
\end{figure}

% Need an intro

%_____SUB-SUB-SECTION____
\subsubsection{Complexity}
The functions affected by code suggestions in our sample show a  median LOC number below 35 across all languages and across all projects.
Figure \ref{fig:fucntion_LOC} summarizes the distribution of the investigated functions in terms of programming language and total Lines Of Code (LOC), excluding comments.
The figure also points out the differences in LOC before and after introducing code suggestions to the original commits.
While this number has remained seemingly constant for functions written in C and Ruby, functions written in Java, Python, or C\# have seen a slight increase, with the median increasing by 3 lines for Java and 2 lines for Python and C\# functions.
An even higher increase is observable for the C++ language where the median LOC has increased from 26 to 33 lines after the introduction of code suggestions.

\begin{figure}[t]
\centering
    \includegraphics[scale=0.45]{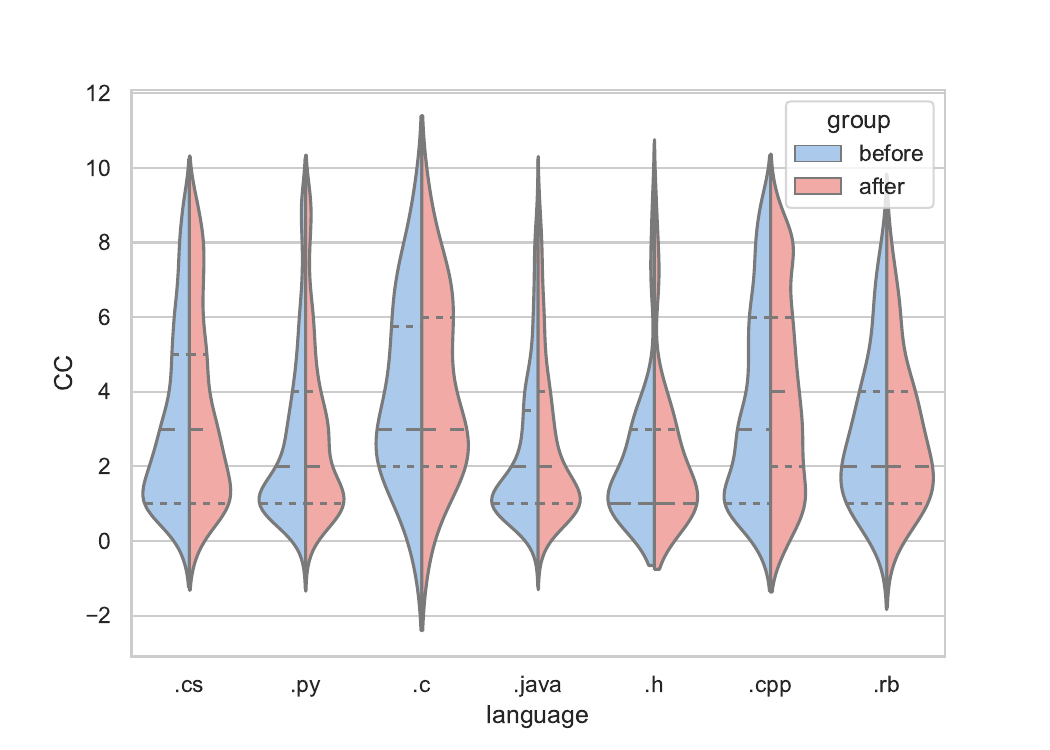}
    \caption{Cyclomatic complexity before \& after the suggestions.}
    \label{fig:function_CC}
\end{figure}

% This analysis helps putting the complexity analysis into context. Relativise
When looking at the difference in cyclomatic complexity of the functions before and after code suggestions a similar trend arises (Figure \ref{fig:function_CC}).
In fact, the distribution of CC remains the same for the C\#, Ruby, and Python before and after introducing suggestions.
Furthermore, the C and Java languages have seen a slight increase in the third quantile to a CC of 4 and 6 respectively, suggesting an elongation of the distribution tail towards the higher CC values.
As is the case for LOC, the highest CC value is observed for C++ functions, where the 25\% quantile and the median have shifted by one point, thus concentrating 50\% of the function in a CC interval of $[2,6]$.
These observations can be explained by the theorised linear relationship between LOC and CC by Graylin et al.~\cite{Complexity_JayHSHKW09}.

As for changes in cognitive complexity described in Table \ref{tab:cognitive_complexity}, we observed that for the majority of functions (92.7\%) this metric remained unchanged.
Functions written in Ruby exhibit the highest change rates in both directions, with an average decrease of 2.67 points affecting 21.4\% of the functions and an average increase of 10 points impacting 3.6\% of the functions.
Notably, the percentage decrease in cognitive complexity exceeds the percentage increase across all programming languages.

% \begin{figure}
%     \includegraphics[scale=0.27]{figures/cc_violin_plot.pdf}
%     \caption{Cognitive Complexity before and after Suggestions}
%     \label{fig:function_CC}
% \end{figure}
% \input{tables/cognitive_complexity_total}

\subsection{RQ3: Impact on Pull Requests}
\label{sec:result_impacts}
%_____SUB-SUB-SECTION____
Overall, the regression models exhibit limited explanatory power, as indicated by the low $R^2$ values. 
Despite this, we proceed to examine and report on the relationships between the predictors and the dependent variables.
\subsubsection{Merge Rate}
The regression analysis for the dependent variable merge rate has shown that the usage of suggestions has a positively significant impact on whether or not a 
pull request gets merged $Y_{\scaleto{MERGE}{3pt}}$.
In fact, $p< 0.05$ for the binary variable \textit{with suggestion}, which also has a positive coefficient $B=0.11$ as outlined in Table \ref{tab:result_stat_merge}.
% Other variables having a significantly positive influence on merge rate include the number of files changed $B= 0.053$.
Interestingly, the submitter's association with the project also significantly influences the outcome with members registering the highest coefficient score $B=0.536$ followed by contributors $B=0.342$ and collaborators $B=0.327$.
% Code churn represented by the number of additions and deletions and number of review comments in a pull request, although significant, have a very low coefficient score.
Lastly, the number of commits and comments  weakly affects the merge decision of pull requests negatively with $B=-0.001$ and $B=-0.004$ respectively.
% \todo{One sentence interpretation}
% \input{tables/result_statistic_merge}

%_____SUB-SUB-SECTION____

\subsubsection{Resolution Time}
The use of code review suggestions significantly influences the time to resolution $Y_{\scaleto{DURATION}{3pt}}$ of a pull request ($p<0.05$) as shown in Table \ref{tab:result_stat_merge}.
In this case, the positive coefficient $B=104.488$ indicates that using suggestions prolongs the time that it takes for a pull request to be resolved.
This variable also registers the highest positive coefficient of all the independent variables, followed by a much lower positive significant coefficient for the total number of comments $B=39.868$ and review comments $B=5.978$.
All other independent variables to the exception of deletions significantly  influenced resolution time negatively.
Here again, the submitter's association with the project plays an important role with members having the pull requests with the fastest resolution time $B=-881.415$ followed by contributors $B=-783.669$ and collaborators $B=-764.321$.
% \todo{One sentence interpretation, more discussions, style wars }
%_____SUB-SUB-SECTION____

\subsubsection{Suggestions and Knowledge Sharing}
We were interested to know if code review suggestions were used beyond the pull request review process as an information source.
We therefore asked the 92 respondents familiar with the feature a multiple choice question about whether or not they have searched code review suggestions from repository history for one or more of the following purposes: 
\begin{itemize}
    \item Find the rationale behind a code section
    \item Learn project coding conventions
    \item Find a suitable reviewer
    \item Find an expert to help with a similar problem
    \item Find an example to implement a code section
    \item Learn how code review is done in the project
\end{itemize}

\begin{figure}
    \centering
    \includegraphics[scale=0.35]{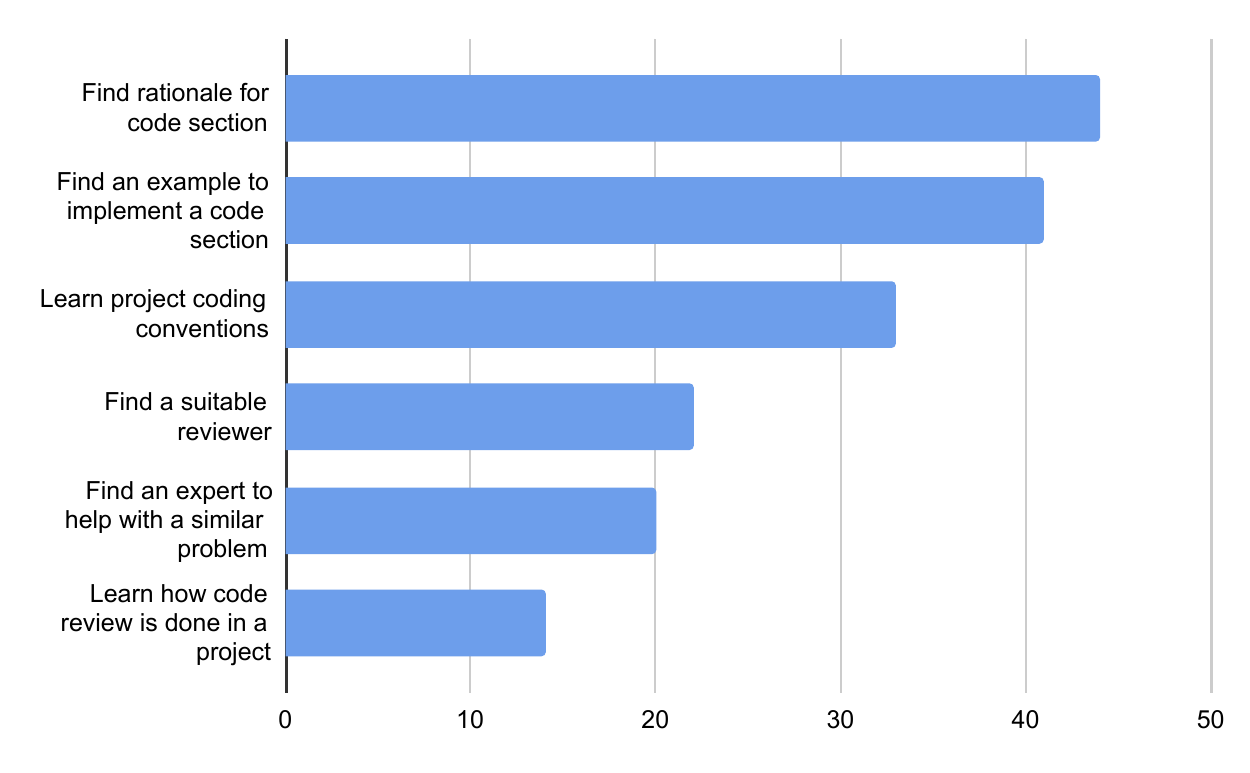}
    \caption{Suggestions as an information source.}
    \label{fig:suggestion_info_source}
\end{figure}

Respondents also had the option of filling in other information as free text.
The results are depicted in Figure \ref{fig:suggestion_info_source}.
Overall, 32 respondents mentioned not searching suggestions at all, one person mentioned using suggestions to \textit{``understand why something has changed''} and one person stated that they rarely skim past PRs unless they are \textit{``looking for discussion around a past change found using blame''}. 
The remaining 58 participants stated searching suggestions for at least one of the information types presented to them, whereby the number of negative responses exceeds the positive ones for each type of the presented information types.
More people stated searching suggestions to find code rationale (44/92) than any other information.
Other popular uses included code reuse whereby developers searched for an example implementation (41/92 positive answers) and learning about project coding conventions (33/92 positive answers).
According to the results, suggestions are less likely to be used when looking for a suitable reviewer, an expert's help or when the goal is to learn how code review is conducted in a project (22/92, 20/92 and 14/92 positive answers respectively).
We conclude that suggestions are used as an information source, mainly when looking for rationale or for a code example  \cite{Sadowski_Search_Code_2015, Widyasari:TOSEM:2024}.

%__________________SUB-SECTION_________________
\begin{table}[t]
\centering
    \setlength{\tabcolsep}{4pt}
    \caption{Changes in cognitive complexity per language.}
    \begin{tabular}{lrrrrr}
        \toprule
        &\multicolumn{2}{c}{\textbf{Increase}}&\multicolumn{2}{c}{\textbf{Decrease}}&\textbf{Unchanged}\\
        \textbf{Lang.}&\textbf{Count}&\textbf{Mean}&\textbf{Count}&\textbf{Mean}&\\
        \midrule
        rb (56)&2 (3.6\%)&10.0&12 (21.4\%)&-2.67&42 (75.0\%)\\
        py (484)&11 (2.3\%)&1.18&21 (4.3\%)&-2.05&452 (93.4\%)\\
        c\# (1223)&21 (1.7\%)&4.76&74 (6.1\%)&-1.66&1128 (92.2\%)\\
        java (441)&9 (2.0\%)&2.0&13 (2.9\%)&-1.54&419 (95\%)\\
        php (48)&1 (2.1\%)&2.0&3 (6.3\%)&-3.67&44 (91.7\%)\\
        scala (10)&0 (0\%)&0&0 (0\%)&0&10 (100\%)\\
        ts (3)&0 (0\%)&0&0 (0\%)&0&3 (100\%)\\
        js (5)&0 (0\%)&0&0 (0\%)&0&5 (100\%)\\
        swift (4)&0 (0\%)&0&0 (0\%)&0&4 (100\%)\\
        \midrule
        \textbf{Total (2274)}&44 (1.9\%)& 3.48 & 123 (5.4\%)& -1.86 & 2107 (92.7\%)\\
        % \textbf{\%}&1.9\%& & 5.4\%&  & 92.7\%\\

        \bottomrule
    \end{tabular}
% \vspace{-1em}
\label{tab:cognitive_complexity}
\end{table}

\subsection{RQ4: Suggestions and Project Roles}
\label{sec:result_impact_roles}
%The perspectives we focused on are the reviewers' motivation for writing the code review suggestion and the role of project experience in the use and acceptance of code review suggestions.
%We used a qualitative approach for the first perspective, stating examples of the different reviewer motivations.
%The investigation of the second perspective was achieved through a quantitative analysis of the coded code review suggestions from section \ref{sec:open_coding}.
To investigate the interplay between the pull request stakeholders' association with the project and the usage of  review suggestions, we first analysed their distribution in our dataset then compared the results to the survey responses.
We asked participants how project roles influence their decision of creating and accepting suggestions.

%_____SUB-SUB-SECTION____

\subsubsection{Usage of Suggestions} 
% In general (ALL)
The results of our analysis based on project association is depicted in Table \ref{tab:roles_sugg_type}.
The table shows that the highest number of suggestions in our coded dataset was shared between members and contributors (807 = 657 + 150) and in particular in cases where the submitter of a pull request is a \textit{Contributor} and the reviewer is a \textit{Member} (657).
The second highest code suggestion activity is registered among contributors (426), followed closely by suggestions among members (392).
In contrast, suggestions from and to developers with no prior project association, i.e. holding the role \textit{None}, are rare.
Developers holding this role exchanged suggestions with members more than any other fraction of developers.
Submitters with the role \textit{None}  received 54 suggestions from members and reviewers of this role made 36 suggestions to members authoring pull requests.
% The lowest suggestion activity was registered for developers having the role \textit{Collaborator}.

The interaction results can be explained by the well-known core-periphery network  \cite{El_Asri_Peripheral_2017}.
The relatively low interaction rate from and to developers having no prior association with the project can be explained by the fact that developers become \textit{Contributors}, as soon as they make their first contribution.
Lastly, the distinct role of \textit{Collaborator} is not used in all projects, which makes developers holding this role and consequently their interactions rare.

%_____SUB-SUB-SECTION____

\subsubsection{Suggestion Type Distribution}
The distribution of the suggestion types among the prolific submitter-reviewer roles mirrors that of the overall distribution in our sample.
In fact, improvements prove to be the most prolific type both among members (53.3\%), among contributors (44\%) and between submitting contributors and reviewing members (51.3\%).
Similarly, documentation is the second most prevalent type for all the role constellations in question with the highest value registered among contributors (35.6\%).
The values for code style and fixes are also comparable to the general distribution with the highest value reported for submitting contributors paired with reviewing members (19.3\%).
% Overall, the results show that contributors tend to suggest a higher number of documentation changes among each other while members tend to point out slightly more defects while reviewing the code of contributors.
\begin{table}[t]
    % \centering
    \caption{Regression analysis.}
    \setlength{\tabcolsep}{11pt}
    \begin{tabular}{l rrrr}
    \toprule
        & \multicolumn{2}{c}{$Y_{\scaleto{MERGE}{3pt}}$} & \multicolumn{2}{c}{$Y_{\scaleto{DURATION}{3pt}}$}\\
        \midrule
         \textbf{Predictors}&\textbf{B}&\textbf{p}&\textbf{B}&\textbf{p}\\
         \midrule
       
        % (Constant)&0.538&0.000& 764.638&0.000\\
        % additions&0.000&0.034 & -0.001&0.000\\
        % deletions&0.000&0.217 & 0.003&0.000\\
        % changed\_files&0.053&0.000 & -0.083&0.002\\
        % commits&-0.002&0.000 & -0.420&0.455\\
        % review comments&0.001&0.001 & -10.155&0.000\\
        % total comments&-0.001&0.028 & 9.429&0.000\\
        % member&0.290&0.000 & -466.286&0.000\\
        % contributor&0.083&0.000 & -322.943&0.000\\
        % collaborator&0.116&0.000 & -384.096&0.000\\
        % with suggestions&0.082&0.000& 456.387&0.000\\

        (Constant)&0.288&$<$.001&1112.513&$<$.001\\
    with suggestions&0.11&$<$.001&104.488&0.003\\
    additions&-0.000&0.177&-0.001&$<$.001\\
    deletions&0.000&0.3&0.003&$<$.001\\
    changed files&0.000&$<$.001&-0.025&0.383\\
    commits&-0.001&$<$.001&-2.274&$<$.001\\
    total comments&-0.004&$<$.001&39.868&$<$.001\\
    review comments&0&0.072&5.978&$<$.001\\
    member&0.536&$<$.001&-881.415&$<$.001\\
    contributor&0.342&$<$.001&-783.669&$<$.001\\
    collaborator&0.327&$<$.001&-764.321&$<$.001\\

        \midrule
        % Adjusted $R^2$&&0.062& & 0.017\\
        Adjusted $R^2$&&0.089& & 0.07\\
        \bottomrule
    \end{tabular}
    % \vspace{-1em}
    \label{tab:result_stat_merge}
\end{table}

\begin{table*}[t]
        \caption{Distribution of suggestion types in relation to pull request submitters' and reviewers' roles.}
        \begin{adjustbox}{max width=1.03\textwidth}
        \setlength{\tabcolsep}{2.1pt}
    \begin{tabular}{rl rrrrr| rrrrr| rrrrr| rrrrr}
            \toprule
            &&\multicolumn{20}{c}{\textbf{ Reviewer}}\\
            \cmidrule{3-7}\cmidrule{8-12}\cmidrule{13-17}\cmidrule{18-22}
            &&\multicolumn{5}{c}{Member}&\multicolumn{5}{c}{Collaborator}&\multicolumn{5}{c}{Contributor}&\multicolumn{5}{c}{None}\\
            \cmidrule{3-7}\cmidrule{8-12}\cmidrule{13-17}\cmidrule{18-22}
                 &&\cellcolor{gri}\textbf{All}&\textbf{F}&\textbf{I}&\textbf{CS}&\textbf{D}&\cellcolor{gri}\textbf{All}&\textbf{F}&\textbf{I}&\textbf{CS}&\textbf{D}&\cellcolor{gri}\textbf{All}&\textbf{F}&\textbf{I}&\textbf{CS}&\textbf{D}&\cellcolor{gri}\textbf{All}&\textbf{F}&\textbf{I}&\textbf{CS}&\textbf{D}\\
                 \midrule
                 
        \parbox[t]{2mm}{\multirow{4}{*}{\rotatebox[origin=c]{90}{\textbf{ Submitter}}}}&Member &\cellcolor{gri} 392 &5.9\%& 53.3\% & 18.9\% & 21.9\% & \cellcolor{gri}4 &0 & 75\% & 25\% &0 & \cellcolor{gri}150 & 6.7\% & 46.7\% & 16.7\% & 30\% & \cellcolor{gri}36 & 16.7\% & 33.3\% & 11.1\% & 38.9\%\\
        &Collaborator&\cellcolor{gri}3&0 & 66.7\% &0 & 33.3\% & \cellcolor{gri}0 &0 &0 &0 &0 & \cellcolor{gri}0 &0 &0 &0 &0 & \cellcolor{gri}4 &0 &0 &0 & 100\%\\
        &Contributor&\cellcolor{gri}657&9.3\% & 51.3\% & 19.3\% & 20.1\% & \cellcolor{gri}9 & 11.1\% & 66.7\% & 11.1\% & 11.1\% & \cellcolor{gri}416 & 7.5\% & 44\% & 13\% & 35.6\% & \cellcolor{gri}14 & 7.1\% & 50\% & 7.1\% & 35.7\%\\
        &None&\cellcolor{gri}54&11.1\% & 38.9\% & 11.1\% & 38.9\% & \cellcolor{gri}2 &0 &0 &0 & 100\% &\cellcolor{gri} 24 & 4.2\% & 20.8\% & 37.5\% & 37.5\% &\cellcolor{gri} 10 & 20\% & 60\% & 20\% &0\\
        \bottomrule
    \end{tabular}
\end{adjustbox}

\begin{flushleft} 
$\quad\quad$ \textbf{F}= Fix, \textbf{I}= Improvement, \textbf{CS}= Code Style, \textbf{D}=Documentation 
\end{flushleft} 

\label{tab:roles_sugg_type}
\end{table*}

%_____SUB-SUB-SECTION____

\subsubsection{Acceptance}
\begin{figure*}[t]
\centering
\begin{subfigure}[b]{0.4\textwidth}
    \includegraphics[scale=0.35]{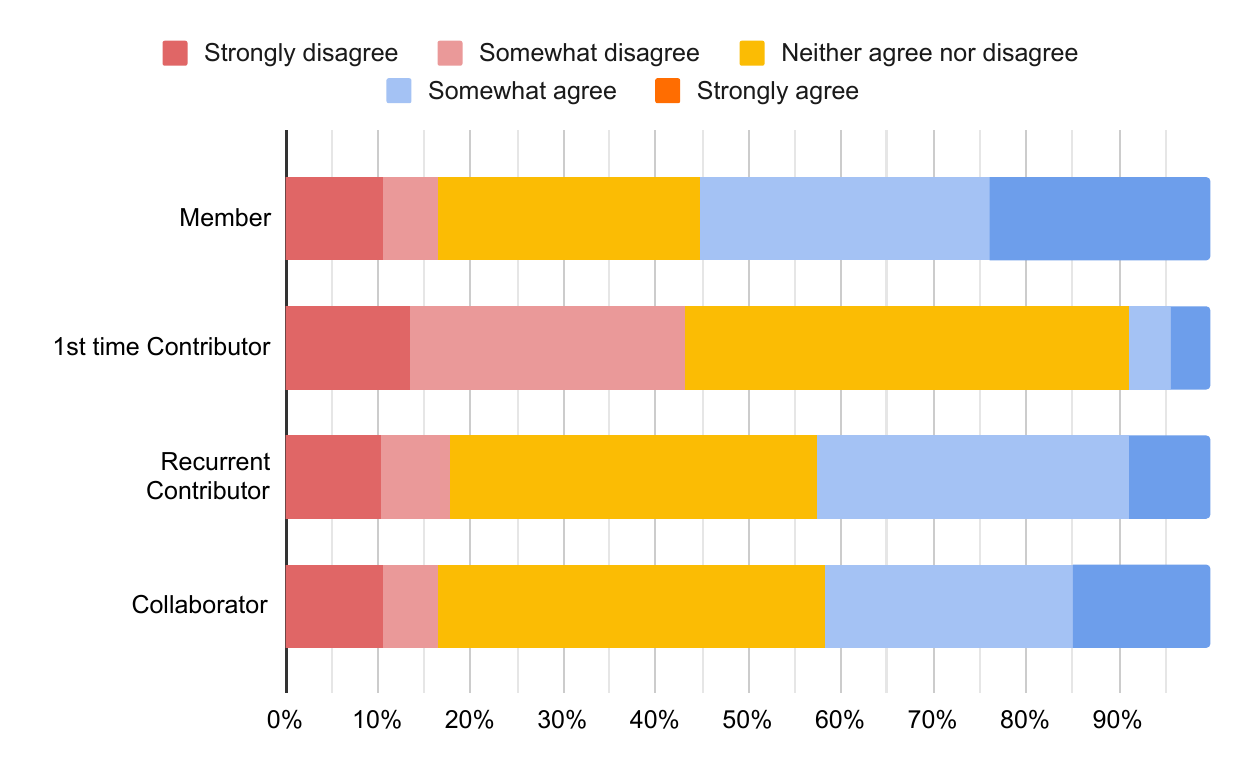}
    \caption{Perception of submitters on the impact of roles.}
    \label{fig:submitter_perception}
\end{subfigure}
\hspace{2em}
\begin{subfigure}[b]{0.4\textwidth}
    \includegraphics[scale=0.35]{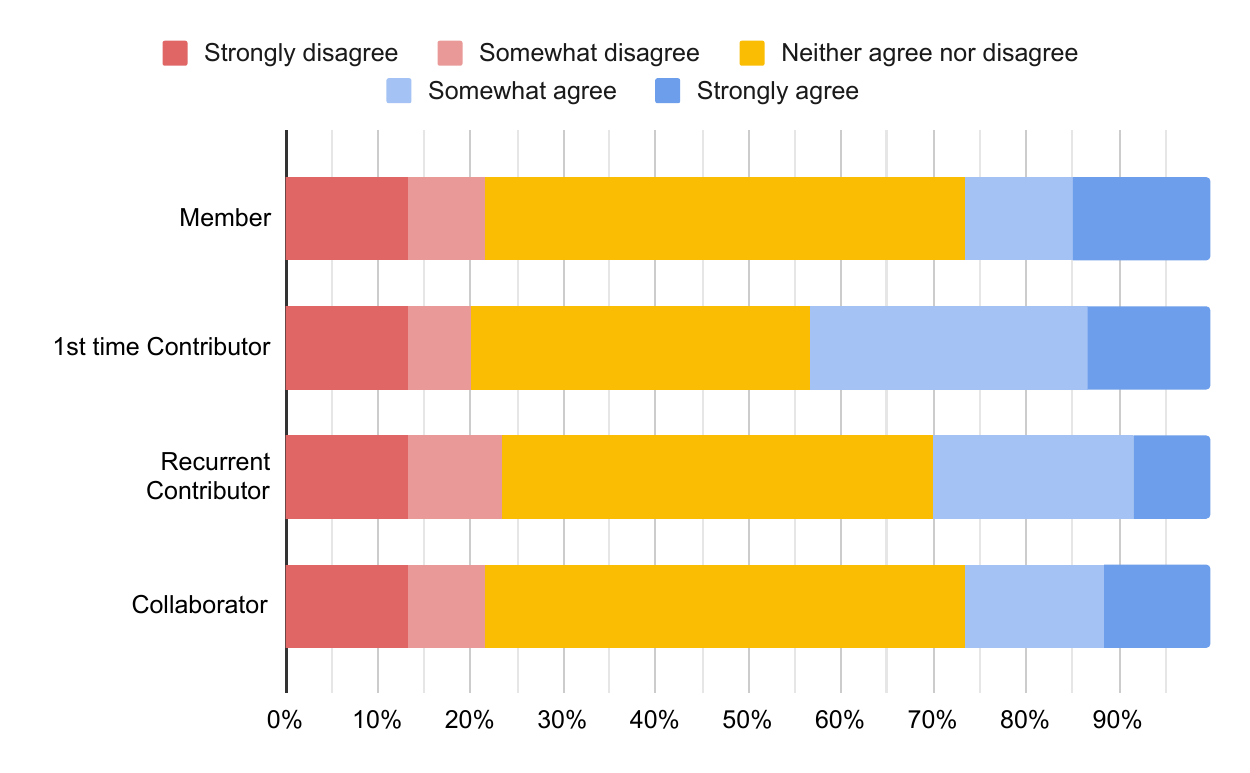}
    \caption{Perception of reviewers on the impact of roles.}
    \label{fig:survey_data}
\end{subfigure}
\caption{Influence of Project Association on Suggestion Usage}
\label{fig:roles2}
\end{figure*}

We asked the participants in our survey a Likert scale question about the influence of user roles on their decisions with regard to suggestions.
The results are outlined in Figure \ref{fig:roles2}.
Here again we used the distinction based on the feature familiarity question to differentiate between submitter and reviewer perspectives.
When submitters were asked if they are more likely to accept a code suggestion from a member, collaborator, recurrent contributor, or first-time contributor,
they gave the highest overall agreement (54.4\%) to suggestions from members while first-time contributors had the least chances of getting their suggestion integrated (8.8\% agreement).
Similarly, first-time contributors had the highest level of overall disagreement (44.1\%) and the highest rate of neutrality (47.1\%).

Reviewers on the other hand were asked if they were more likely to use suggestions if the pull request submitter is a member, collaborator, recurrent contributor, or \textit{first-time contributor}.
Reviewers were most likely to use suggestions when the pull request submitter is a first-time contributor as attested by the highest agreement score (44.1\% slight to strong agreement).
In terms of agreement score, recurrent contributors were second most likely to receive suggestions (28.8\%) followed by collaborators (25.4\%) and members (25.4\%).
Notably, the rate of neutral responses is the lowest for fisrt-time contributors (35.6\% neutral response) compared to the rest of the roles, where it reached as high as 50.8\% for members and collaborators.
Disagreement scores on the other hand did not differ significantly between the roles.

%--------------------------SECTION---------------------------

\section{Discussion}
\label{sec:discussion}

Our findings reveal that \textbf{improvement and documentation} suggestions are more prevalent than fix and code style suggestions, aligning with the goal of maintainability, which is highly prioritized in the code review context. 
The relatively small number of code style suggestions can be attributed to the widespread adoption of linters and continuous integration tools that automatically address formatting issues and reduce code defects. 
We observed that code review suggestions often involve \textbf{small code changes}, which positively impact merge rates but can sometimes delay pull request resolution due to discussions. 
The usage of suggestions varies by stakeholder roles, with members focusing more on defects, likely due to differences in project experience and better knowledge of change impact. 
Furthermore, we found that code review suggestions serve as valuable \textbf{educational} tools, guiding newcomers, and supporting knowledge sharing by serving as a searchable or recommendable information source \cite{Widyasari:TOSEM:2024}.

% \subsection{Developer Reception and Implications}
As the code suggestion and similar structured feedback features gain maturity, we can expect further studies to investigate it from different perspectives and for different usage scenarios.
This is already true for SAW-BOT \cite{SAW-BOT_2021}, a GitHub bot proposing fixes to static analysis warnings using code review suggestions.
% The bot developed by Dragos et al. leverages the benefits of suggestions and their level of acceptance by GitHub developers as outlined by the authors' interview results.
% However, it remains unclear how the reception among developers for such solutions  will be especially since bots have a reputation for spamming.
Even though the use of bots is on the rise on GitHub, they have a mixed record among developers. 
Wessel et al.~\cite{Wessel_GitHub_Bots_2018} reported that a third of the projects they investigated used at least one type of bots.
However, the authors report 16 challenges developers face with bots, most prominently their poor decision support.

Mirroring Wessel et al.'s work, we asked the participants of our survey an optional open-ended question on what they think about the automation of code review suggestions. 
We also ask them to state a scenario where such an automation can reduce the code review effort in their opinion.
The majority of respondents said that an automation can prove useful but only when restricted to \textbf{simple use cases} such as code style, formatting and typos. 
Others said that even these problems should be left to already existent tools such as linters and some suggested integrating linter checks with code review suggestions.
On a similar note, some respondents saw it useful to integrate static analysis within code review suggestions: \textit{``Having findings from something like coverity be added to PRs automatically as code review suggestions could be quite useful.''}
Another use case mentioned by two respondents is the explanation of CI failures: \textit{``Maybe automated code review comments could explain how to deal with project-specific CI failures.''.}
Finally, one respondent suggested creating \textit{``code recommendation based on similarity with other parts of the project''}.
Furthermore, concerns were raised about the intrusiveness  of bots as in the following responses: \textit{``Bots can be spammy''},\textit{`` they would be rather annoying if they are too extensive or trivial''}, \textit{``[...] it could easily be off-putting and spammy if a bot quickly creates hundreds of suggestions.''}
Others were also concerned about the precision of a potential automation stating that they are \textit{`` prone to false-positives, which make automated code review suggestions less useful on average''}, another participant mentioned : \textit{`` I could not stand suggestions that would be flawed even 10\% of the time''}.

Based on the results from our study and the need for ``smarter bots'' outlined in the work of Wessel et al.~\cite{Wessel_GitHub_Bots_2018}, we recommend integrating context-awareness for automation efforts using code review suggestions \cite{Maalej:PAKM:2008, maalej2023rsse}.
Such efforts should leverage project and developer attributes to provide targeted support and avoid spamming code review discussions.
Project characteristics may include coding conventions and project history, particularly changes made to previous pull requests and their acceptance rates.
Developer characteristics may include their roles, project experience, and  past activity within the project such as reviewer's most recurrent suggestions.
Automated suggestions can also be targeted for newcomers by providing alternative implementation examples.

% To gain the developers' opinion on this subject, we included an optional free text question in our survey asking developers if they could think of an example in which automating code review suggestions might be useful.
% The overwhelming majority of the responses we received mentioned that such an automation only makes sense for small code style and ``cosmetic'' changes.
% A considerable number of respondents voiced concern about the rate of possible false positives "".
% Others mentioned that such bots would add to the amount of annoying spam messages received by Github bots.

\section{Threats to Validity}\label{threasts}
% \textbf{Conclusion Validity.} 
\textit{\textbf{Internal validity.}} 
Open coding of suggestion types was sometimes challenging due to a lack of change context, making it unclear if changes were improvements or bug fixes. To avoid misjudging severity, we relied on natural language comments but still recorded 557 coding mismatches. Most conflicts (213/557) were between improvements and code style suggestions (Mismatch Type I), followed by improvements vs. fixes (154/557, Type II) and improvements vs. documentation (97/557, Type III), comprising 83.3\% of all conflicts. We reviewed 30 sample comments per type to analyze sources: Type I often stemmed from code style inconsistencies, Type II from ambiguous descriptions of change severity, and Type III from suggestions affecting program output (e.g., log/exception changes). For reliability, we focused on the agreement set from open coding.
To reduce participant selection bias, we invited all contributors from the studied projects to participate in our survey. However, we could only reach contributors who publicly shared their email and name, limiting our sample from over 10,000 to 3,771 GitHub users. We only contacted users who had made contributions, excluding potential contributors without documented activity. Indicators like stargazing, watching, or forking were not considered reliable signs of contribution intent. Another internal validity concern is that our responses may primarily reflect users interested in code review in open-source projects.
\textit{\textbf{Construct validity.}} 
% When asking about their familiarity with the suggestion feature, we provided the participants with an image exemplifying a GitHub suggestion in order to prevent misinterpretation of terminology.
% Furthermore, we tested the reliability of our chosen 5 point Likert scale by calculating Cronbach's alpha \cite{Gliem_Cronbach_2003} for each question and participant group.
% For the Likert scale question on code review suggestion types, $\alpha=0.827$ and $\alpha=0.661$ for pull request submitters and reviewers respectively, which is considered acceptable \cite{Truong_Crohnbach_2020}.
% For the Likert scale question on the influence of user roles on the use and acceptance of code review suggestions, $\alpha=0.874$ and $\alpha=0.901$ for pull request submitters and reviewers respectively.
% Because a participant could have experience with suggestions both as pull request submitter and reviewer, we tried to formulate the questions differently for each perspective to avoid redundancy and learning bias.
%%%%%%%
To ensure clarity on the suggestion feature, we provided participants with an example image. We assessed the reliability of our 5-point Likert scale using Cronbach’s alpha, finding acceptable scores: $\alpha = 0.827$ for suggestion types, and $\alpha = 0.874$ and $\alpha = 0.901$ for the influence of user roles on suggestion use, for submitters and reviewers, respectively. To minimize redundancy and learning bias, we tailored questions based on participant roles.
\textbf{External Validity.} 
% Even though our project sample may not be representative of the general population of projects on GitHub, our aim has been to study engineered open source projects in particular.
% We furthermore used thresholds known from literature, which may have the side effect of excluding smaller projects \cite{Commit_Classification_Levin_2017}\cite{Repo_Dataset_Munaiah_2017}.
% Accordingly, the participants in our sample may not be representative of the suggestion feature's user base on GitHub.
Although our sample may not represent all GitHub projects, our focus was on engineered open-source projects, using literature-based thresholds that may exclude smaller projects. Thus, our participants may not fully represent all users of the suggestion feature on GitHub.

\section{Related Work}
\label{sec:relwork}

The work of Brown and Parnin \cite{Brown_Suggestions_2020} was among the first to present an empirical investigation the suggestion feature's usage, its effectiveness, and the perception of its usefulness by developers.
The authors analysed a total of 17,712 suggestions from 22 GitHub projects and surveyed 43 developers.
They summarised the benefits of integrating suggestions in five main aspects: timing, location, actionability, conciseness and communication.
% While suggestions potentially lengthened the code review process, the developers in the study commented that they were able to make recommendations much faster making for a better timing.
% Furthermore, developers did not need to leave the pull request page to make a suggestion, which covers the aspect of locality.
% Actionability on the other hand is given by the ability to incorporate the change immediately simply by committing the suggestion while conciseness was explained by the fact that developers prefer code over natural language recommendations.
Our categorisation of suggestion types shows strong parallels to their work  with subtle differences in  terminology.
The authors namely referred to \textit{documentation} suggestions as \textit{non-functional} suggestions and to suggestions proposing a \textit{fix} as \textit{corrective} suggestions.
Moreover, the fourth category in their work is referred to as \textit{formatting}, which in our case represents a subset of the type \textit{code style}.
The distribution of suggestion types in our results partly confirms the results reported by the authors \cite{Brown_Suggestions_2020}.
Both studies found that suggestions of the types \textit{improvement} and \textit{documentation}  (\textit{non-functional}) occur more  frequently than suggestions of the types \textit{fix} (\textit{corrective}) and \textit{code style} (comparable with \textit{formatting}).
However, while Brown \& Parnin's work shows a similar occurrence frequency for \textit{non-functional} and \textit{improvement} suggestions (36\% and 34\% respectively), \textit{improvements} surpass \textit{documentation} suggestions in our assessment.
% Similarly, the gap in occurrence count between \textit{corrective} and \textit{formatting} suggestions is less prominent in Brown et al.'s work (16\% and 14\% respectively) than it is between \textit{code style} suggestions and \textit{fixes} in our sample.
This discrepancy may be explained  by the difference in sample sizes: our sample contained 1,775 manually labelled examples drawn from a population of 8,672 suggestions, while Brown \& Parnin used a random sample of 100 examples from 17,712 suggestions.
The perception of usefulness for the different types is comparable with the authors' results on the overall feature usefulness from both stakeholder perspectives.

% Our impact analysis further unveiled that suggestions mainly consist of small code changes, which did not lead to a significant decrease in CC and cognitive complexity, despite simplification being a recurrent motivation behind improvement suggestions. 
% While this is partly explained by the linear relationship of CC and LOC, we recognise that other complexity metrics such as cognitive complexity \cite{Campbell_Cognitive_Complexity_2018} might be more adequate in conveying the human perception of complexity.
Our impact analysis showed that the usage of suggestions positively affected the merge rate and significantly prolonged the pull request's resolution time.
Our results concur with those presented by Brown \& Parnin \cite{Brown_Suggestions_2020} on resolution time but not on merge rate.
One possible interpretation of these results is that suggestions demand effort on the reviewers' part.
A reviewer might be more likely to invest time and write a code review suggestion if they already agree with the pull request and recognise its necessity. 
Following this assumption, a higher merge rate may simply have co-occurred with the use of suggestions without necessarily inferring a causal relationship.
Although suggestions can be used to optimise the code review process through shorter iterations, they can also spark discussions delaying the resolution.
For instance, a participant in our study mentioned the drawback of \textit{``getting caught up in style wars''}.
In fact, conflicts may arise during code review.
In an interview study with OSS developers, Wurzel Gon{\c{c}}alves et al.~\cite{wurzel2022interpersonal} found that code readability and style issues were the main non-functional sources of conflict.

% The prevalence of improvement and documentation suggestions is en par with the goal of maintainability, which was ranked the most important code review goal in a survey with  open source developers presented by Bosu et al. \cite{BosuCodeReviewOSSMicrosoft}.
% The relatively small number of code style suggestions on the other hand can be attributed to the use of linters which have become widely adopted by many projects.
% In addition, the widespread adoption of continuous integration tools  could have contributed to the reduction of code defects, which explains the small number of fix suggestions.

Generally, according to a knowledge sharing measure defined in the work of Rigby et al.~\cite{Rigby_Convergent_Contemporary_CR_2013}, conducting peer reviews increases the number of distinct files known to a developer by  66\%. 
The results of our survey have shown that code review suggestions also contribute to knowledge sharing by acting as a searchable information source, primarily for code rationale—often stated as a rare commodity \cite{Dutoit:Book:2006, LaToza_Questions_Code_2010, Maalej:TSE:2013}.
Code suggestions were also consulted when looking for code examples, which were mentioned by Sadowski et al.~\cite{Sadowski_Search_Code_2015} as being  in high demand among developers.
Rationale and code examples both figure among the categories of explanations presented by Widyasari et al.~\cite{Widyasari:TOSEM:2024} in their taxonomy of explanations in code review comments. 
The authors recently studied the different types of explanations commonly used in Gerrit code review comments and categorized them into seven types, each reflecting how reviewers communicate feedback and attempt to clarify their reasoning to pull requests authors.
% These findings also confirm that code review suggestions contribute to knowledge sharing, which ranked the second most important code review goal by Bosu et al. \cite{BosuCodeReviewOSSMicrosoft}.
% We should note however that these differences may be due to the differences of magnitude between our sample and that of the authors, whose sample consisted of 100 randomly selected suggestions from a larger population of 17,712 suggestions.
% Moreover, the prevalence of improvements in our dataset may be partly explained by our strict definition of code style suggestions in terms of code structure, formatting, commenting and naming.
% In fact, this definition relegates all suggestions affecting source code and not treating any of the defined code style aspects to the improvement category.
% Knowledge sharing is stated by Bosu et al. \cite{BosuCodeReviewOSSMicrosoft} as the second most important goal of the code review process in OSS.
% We chose to further investigate the aspect of communication in our work by perceiving code suggestions as a special type of interaction within the flow of the pull request discussions.

Pull request discussions play a a central role in the development and evolution of open source projects as a whole as outlined by the works of Tsay et al.~\cite{Tsay_Discussions_GitHub_2014}, \cite{Tsay_Social_Factors_2014}.
Developer discussions are commonly guided by technical and social factors and are used beyond quality assurance as a means of building a community around the project.
Bosu et al.~\cite{BosuCodeReviewOSSMicrosoft} further found that code review is used in OSS to form impressions about teammates, which impact future collaborations.
In our study, we showed that the usage of suggestions differs depending on the association of the pull request stakeholders with the project.
Unsurprisingly, we found that the highest usage of code review suggestions was registered between submitting contributors and reviewing members and that the second highest suggestion activity was registered among contributors.
However, we discovered subtle differences between stakeholder roles when looking at the distribution of suggestion types.
In fact, our analysis has shown that contributors tend to use more documentation suggestions  while members tend to point out slightly more defects. 
This can be explained by the differences in project experience, making members more aware of the implications of certain changes.

% Project experience also plays a role in the reviewer's decision on whether or not to use suggestions in their review and in the pull request submitter's decision on whether or not to accept a review suggestion.
In our survey, reviewers indicated they were more inclined to use suggestions when the pull request was submitted by a newcomer to the project.
These findings align with Bosu et al.~\cite{BosuCodeReviewOSSMicrosoft}, who observed that reviewers in OSS sometimes adopted a mentoring role when the author lacked project or programming experience. 
Additionally, our study reveals that reviewers often provide code examples or reference well-written code  when contributions were of low quality. 
These insights suggest that code review serves as an educational tool, where experienced developers guide and support newcomers.

Finally, Palvannan and Brown \cite{palvannan2023suggestion} have explored leveraging GitHub’s suggestion feature to automate code reviews.
They introduced SUGGESTION BOT, which proposes code edits to pull request authors through code suggestions.
The study demonstrated that the bot not only reduced pull request turnaround time but also provided clear, actionable feedback, underscoring its potential to enhance human-bot collaboration in software development.
Recently, Maalej~\cite{maalej2023rsse} argued that even though bots in software engineering (BotSE) have evolved from simple task automators to intelligent assistants, their primary focus remains on information access rather than knowledge sharing among stakeholders. 
The author advocates for further research into bots that capture design rationale or mediate pair programming conversations, emphasizing their untapped potential to enhance knowledge exchange by connecting developers with experts and facilitating the documentation of valuable insights. 
This underscores the need for greater socio-technical context awareness in future research.

\section{Conclusion}
\label{sec:conclusion}
% In this paper, we presented a multi-faceted empirical investigation of code review suggestions in pull requests.
% We first studied the content of suggestions and were able to uncover four types, for which we gathered developer perception from both the perspective of pull request submitter and the reviewer.
% We also studied the impact the usage of suggestions has on pull requests.
% We found that while using suggestions increased the chances of a pull request to be merged, it significantly increased the resolution time and did not decrease code complexity.
% We furthermore found that the usage and acceptance of suggestions varied based on the stakeholders' association with the project with experienced members using suggestions as an educational tool for guiding project newcomers.
% Lastly, we provided recommendations for future automation efforts of code review suggestions based on the feedback provided  by our survey participants.
%%%%%%%%
This paper presents a multi-method investigation of code review suggestions in pull requests. 
We identified four suggestion types and examined developer perceptions from both submitter and reviewer perspectives. 
Our findings show that while suggestions increase the likelihood of pull requests being merged, they also extend resolution time and do not reduce code complexity. Usage and acceptance of suggestions seem to vary by project role, with experienced members often using them as an educative or knowledge sharing tool for newcomers. Finally, we offer recommendations for future automation of code review suggestions based on participant feedback and related work.

% Potential implications: 
% How can we get more developers to suggest improvements more efficiently?
% Can we automate code suggestions? 
% Suggestion bots? 

% Can our analysis be applied for single projects? What use cases could emerge?  

\section{Acknowledgments}
 We thank the survey participants for their support. This work was partly funded by the Deutsche Forschungsgemeinschaft (DFG German Research Foundation), Project NR 166725071.

\balance
\bibliographystyle{IEEEtran}

\bibliography{bibfile}

\end{document}